\documentstyle[11pt,newpasp,twoside,epsf]{article}
\def\lnls {$log\,N-\,log\,S$}
\def\fxfopt     {$F_x/F_{opt}$}

\def\lax    {${_<\atop^{\sim}}$}
\def\alpe {$\alpha_E$}

\reversemarginpar

\begin{document}
\title{The Chandra Multi-wavelength Project (ChaMP): a serendipitous
survey with Chandra archival data.}
 \author{B. J. Wilkes, P. Green, R. Brissenden, R. Cameron, A. Dobrzycki, J. Drake,
N. Evans, A. Fruscione, T. Gaetz, M. Garcia, H. Ghosh,
J. Grimes, J. Grindlay, E. Hooper, M. Karovska, V. Kashyap, D.-W. Kim,
K. Kowal, H. Marshall, A. Mossman, D. Morris, J. Nichols, A.
Szentgyorgyi, H. Tananbaum,
L. van Speybroeck, A. Vikhlinin, S. Virani, P. Zhao}
 \affil{Smithsonian Astrophysical Observatory (SAO), 60 Garden St., Cambridge,
MA 02138, USA}
 \author{J. Baldwin, A. Kindt (MSU), F. Chaffee (Keck), A. Dey, B.
Jannuzi (NOAO), C. Foltz (MMTO), S. Mathur (OSU), B. McNamara (Ohio),
H. Newberg (RPI)}

 \begin{abstract}
         The launch of the Chandra X-ray Observatory in July 1999 opened a
new era in X-ray astronomy. Its unprecedented, $<0.5$" spatial resolution and
low background are providing views of the X-ray sky 10-100 times fainter 
than previously possible. We have initiated a serendipitous survey
(ChaMP) using Chandra archival data to flux limits covering the 
range between those reached by current satellites and those of the small area 
Chandra deep surveys. We estimate the survey will cover $\sim 5$ sq.deg. per
year to X-ray fluxes (2$-$10 keV) in the range 10$^{-13}-6 \times 10^{-16}$
erg cm$^{-2}$ s$^{-1}$
discovering $\sim$2000 new X-ray sources, $\sim 80$\%
of which are expected to be active galactic nuclei (AGN). 
The ChaMP has two parts, the extragalactic survey (ChaMP) and
the galactic plane survey (ChaMPlane).
Over five years of Chandra operations, the ChaMP will provide both a 
major resource for Chandra observers and a key research tool for the study of
the cosmic X-ray background (CXRB) and the individual source populations which 
comprise it.  ChaMP promises profoundly new science return on a number of  
key questions at the current frontier of many areas of astronomy including
(1) locating and studying
high redshift clusters and so constraining cosmological parameters,
(2) defining the true population of AGN, including
those that are absorbed, and so
constraining the accretion history of the universe, (3) filling in the gap
in the luminosity/redshift plane between Chandra deep and previous surveys
in studying the CXRB, (4) studying
coronal emission from late-type stars as their cores become fully convective
and (5) search for cataclysmic variables (CVs) and quiescent Low-Mass
X-ray Binaries (qLXMBs) to measure their luminosity functions.

In this paper we summarize the status, predictions and initial results
from the X-ray analysis and optical imaging.

\end{abstract}

 \section{Introduction}
The large fields-of-view (FoVs) typical of X-ray imaging instruments have 
long been exploited for serendipitous surveys, resulting in far-reaching
and fundamental advances in our knowledge of the X-ray universe and indeed
the universe as a whole ({\it e.g.},
the Einstein Medium Sensitivity Survey (EMSS,
Stocke et al. 1991, Maccacaro et al. 1982);
the Cambridge-Cambridge ROSAT Serendipitous Survey (CCRSS,
Boyle, Wilkes \& Elvis 1997);
ROSAT International X-ray/Optical Survey (RIXOS, Page et al. 1996)).
The next decade promises to be rich for X-ray
astronomy as two major observatories, Chandra and XMM, 
open new vistas with their
deeper flux limits and superior spatial and spectral resolution. 

\begin{figure}[h]
\plotfiddle{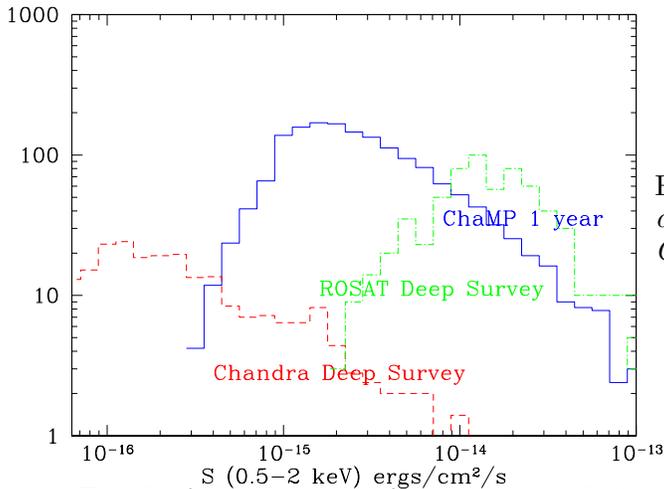}{2.5in}{-90}{33}{33}{-200}{210}
\vskip -2.0in
\hskip 3.0in
\begin{minipage}{2.5in}
\caption{\it Predicted source counts per flux bin for ChaMP Cycle 1.\\}
\end{minipage}
\vskip 0.85in
\end{figure}

To take full advantage of the rich serendipitous dataset afforded by
Chandra's large FoVs (ACIS-I: 16$'\times 16'$)
and to maximize the science per ksec of observing time, we have
organized an extragalactic 
serendipitous X-ray survey, the 
ChaMP and a similar Galactic plane survey (ChaMPlane).
We estimate a ChaMP X-ray sample of $\sim\,2000$ 
sources in the $\sim\,5$ sq. degs. of ChaMP fields in Cycle-1.
$\sim 80$\% of these are likely to be AGN
of various types, including those with low luminosity and/or substantial
low energy X-ray absorption expected based on the spectrum of
the CXRB and already being found in early results from Chandra surveys
(Hornschemeier et al. 2000, Mushotzky et al. 2000,
Barger et al. 2000, Giacconi et al. 2000).
The remaining sources will include clusters of galaxies, 
stars, X-ray binaries, and supernova remnants. Unlike targeted, small area,
deep Chandra surveys, which reach fainter flux levels, the ChaMP provides
the wide
area essential for finding rare, bright and/or unexpected sources, generating
statistically meaningful samples of rare source types such as BL Lac objects,
quiescent X-ray binaries and high-redshift clusters. As such the ChaMP
complements existing deep Chandra surveys (Figure~1).

\section{ChaMP:Data Analysis}
\subsection{Field List: ChaMP, extragalactic survey}

Of prime importance in carrying out any survey is an accurate
understanding of the selection effects present. We
have defined a subset of Chandra fields to minimize bias
due to the effects of known extended sources and to include a variety of
target types.
Each field will be analyzed as it becomes available in the archive
using uniform X-ray source detection and analysis techniques
(Section~2.2)
resulting in a well-defined X-ray sample over the full survey area.

The 172 fields in the ChaMP extragalactic Cycle-1 and Cycle-2 lists
have $|b|>20^{\circ}$.
We exclude targeted survey fields, {\it e.g.} Hubble Deep Field,
Lockman Hole (being
studied optically by their Chandra PIs);
special instrument modes ({\it e.g.}, grating observations,
ACIS sub-arrays); unstable detector periods;
$< 4$ ACIS chips and fields containing bright optical or
X-ray sources covering $>10$\% of the Chandra FoV on the DSS optical or
ROSAT/other X-ray images.

 \subsection{X-ray Data Analysis}
\label{sec:xray}

We are currently refining our source detection procedures using 
two methods: sliding cell search and wavelet transformation; 
the former is optimized for point source detection while the latter
provides efficient detection of diffuse/extended sources
and those in crowded fields. The Chandra point-spread-function (PSF)
is a strong function of off-axis angle and energy so we are
carrying out simulations of source detection efficiency as a function of
size, off-axis angle, background characteristics, source spectrum etc.
to understand survey limits and minimize false sources while
optimizing source detection. 
X-ray source properties will be derived for sources in the master (merged) list
including: accurate positions, 
count rates and fluxes, spatial extent, variability and spectrum/hardness
ratio when possible given the source signal.
ACIS sources with 100 
counts will constrain simple power law fits to within
25\%, 30 counts provide hardness ratios adequate to
distinguish between the high energy spectra of AGN and typically
softer stellar sources and to provide an estimate of the
soft X-ray absorption. Our \lnls\ estimates suggest
$\sim 400$ AGN per year will yield useful hardness ratios, and $\sim 130$
will furnish enough counts for a spectrum.


\begin{figure}[h]
\plotfiddle{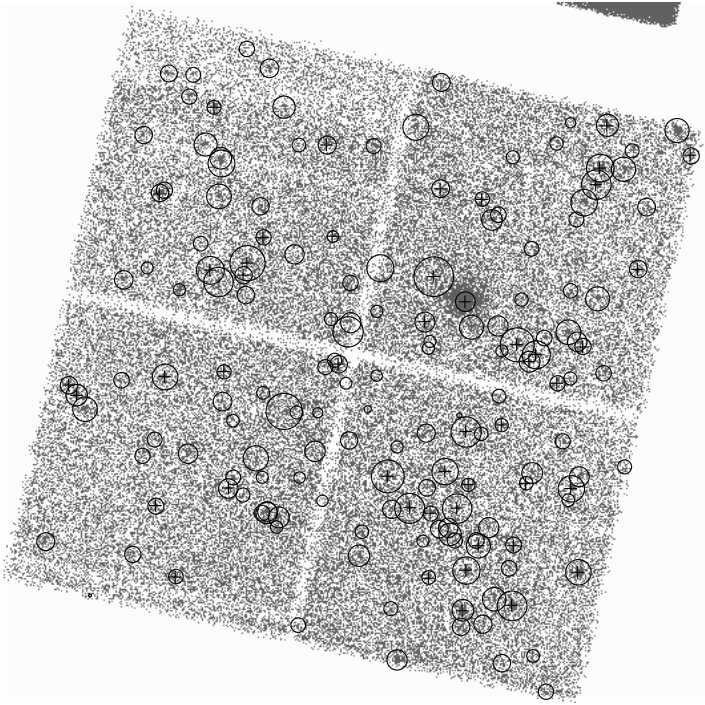}{3.0in}{0}{50}{50}{-210}{-100}
\vskip -2.0in
\hskip 3.4in
\begin{minipage}{2.05in}
\caption{\it ACIS data for the MS1137.5+6625 cluster showing the 
X-ray sources (circles, size $\propto$ log(significance)).
Crosses indicate 
48 possible optical counterparts $<$2 \arcsec.
}
\end{minipage}
\vskip 0.6in
\end{figure}

 \subsection{Optical Imaging}
The second essential ingredient for an X-ray survey, which opens up a much
wider range of scientific questions,
is optical identification of the
X-ray sources. We are 
obtaining optical images of the ChaMP fields
using the same filters with matched flux limits: 
\fxfopt\  $< 0.5$ designed to identify $>$90\% of the normal AGN population
and automatically includes larger fractions of other known source
types. This is very efficient, observing all the X-ray sources
in one Chandra field at
one time, minimizing the large, ground-based telescope time
required to reach the faint optical limiting magnitudes (21\lax V\lax
25, with mode of 24).
The choice of filters, Sloan Digital Sky Survey (SDSS) g$'$,r$'$,i$'$,
provides direct benefit from the
SDSS photometric/spectroscopic database and simulations 
at brighter limits, allowing excellent photometric classification
and redshift determination (Vikhlinin et al. 1998),
invaluable at the faintest optical fluxes.
To date we have observed 41 ChaMP fields in 31 observing nights 
(9 nights:4-m KPNO (Kitt Peak National Observatory), 6:0.9m KPNO,
3:4m CTIO (Cerro Tololo Inter-American Observatory), 17:1.2m SAO).
The uniformity of the dataset and our analysis methods will
provide a well-understood and highly uniform sample.

 \subsection{Optical Identification}
Chandra's small PSF and unprecendented, for X-ray telescopes,
astrometric calibration
minimize the number of possible optical counterparts per
X-ray source. 
We are developing procedures to reduce and analyze the optical images,
derive source lists and cross-correlate (solving for aspect differences),
to identify the X-ray sources.
In Figure~2 we show preliminary results
for a 110 ksec ACIS-I observation based on
a preliminary 900 sec i$'$-band image taken with the
4-shooter CCD imager on the
SAO 1.2m telescope, 12 Feb 2000.
The flux limit,  i$' \sim 18$, is $\sim \times 15$ brighter than the ChaMP required
limit. The 
165 detected X-ray sources found using the
CIAO (Chandra Interactive Analysis
of Observations) {\it wavdetect} tool
and those with optical 
identifications $<2$\arcsec of the X-ray centroid.

We will determine upper limits in optical/X-ray as appropriate for
sources not detected in both bands.
Sources will be classified using X-ray size, optical and X-ray colors
employing color-color plane methods (Newberg et al. 1999).

\subsection{ChaMP Archive}
We will release a list of X-ray sources and their properties
along with optical images and source
lists in to a web-based public archive (the X-ray data will
be available via the Chandra public archive). The ChaMP database
will provide an invaluable, multi-wavelength database to  
the community, helping to ensure that the science return from Chandra
reaches its full potential.

\section{ChaMPlane: galactic survey}
ChaMPlane selected 15 fields in Cycle-1 such that $|b|<10^{\circ}$ and exposure
time $> 30$ ksecs. The X-ray data analysis is being carried out in close
collaboration with the ChaMP (Section~2.2). However optical
imaging in R and H$\alpha$ filters is aimed at optimizing the identification of
CVs and qLMXBs, primarily black hole X-ray novae in quiescence, in order to
constrain and ultimately measure their luminosity functions.
Secondary objectives are to determine the Be X-ray binary content
and stellar coronal source distributions in the Galaxy.
The deep Chandra galactic fields will detect 2-6 keV fluxes (allowing
for low energy absorption) of F$_x$(2-6 keV) = 2
$\times 10^{-15}$ erg s${-1}$ and thus CVs or qLMXBs
with L$_x$ = 10$^{31}$ erg s$^{-1}$ out to
$\sim$7 kpc. Thus most CVs and qLMXBs in the Galaxy can be reached,
and the  ChaMPlane survey offers the best
chance for constraining their formation/evolution and
the stellar BH content of the Galaxy.
CVs and qLMXBs will be identified by their ubiquitous H$\alpha$ excess
as ``blue'' objects in the R vs. (H$\alpha$ - R) plane down to
R $\sim$ 24. We have demonstrated this technique for crowded fields
with our HST discovery of the first CVs in globular cluster cores and
have now conducted a successful pilot ChaMPlane survey at CTIO.
ChaMPlane has been awarded 31 4-m nights of observing time at KPNO 
and CTIO as part of the NOAO Survey Program.

 \section{Preliminary Results}

 \subsection{X-ray \lnls }

We have generated a preliminary \lnls\ from the CIAO {\it wavdetect}
list of 165 background X-ray sources in the MS1137.5+6625 field
and the 127 sources in the field of the cluster: CL 0848.6+4453.
Our results, which assume a power law with \alpe=1.0,
(Figure~3) agree well with those from ROSAT
(Hasinger et al. 1998) and extend to fainter fluxes.

\begin{figure}[h]
\plotfiddle{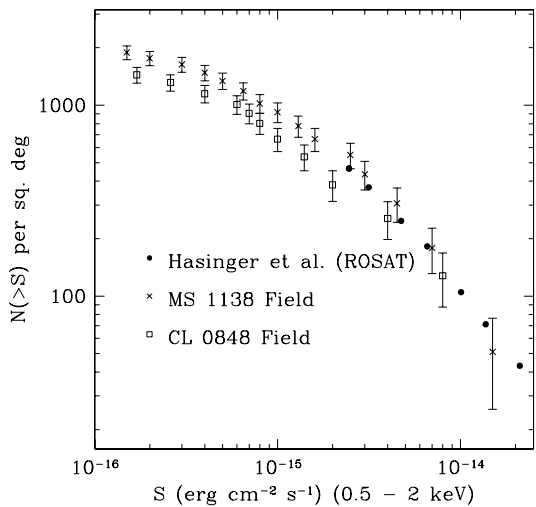}{2.5in}{0}{45}{45}{-220}{-100}
\vskip -2.0in
\hskip 3.4in
\begin{minipage}{2.1in}
\caption{\it Preliminary ChaMP \lnls\ (crosses: MS 1137.5+6625; squares:
CL 0848.6+4453) with 1$\sigma$ error bars, compared
with the ROSAT deep survey (circles, Hasinger et al. 1989).
}
\end{minipage}
\vskip 0.5in
\end{figure}

 \subsection{Optical Spectroscopy}
We plan to take advantage of the new generation of large area,
multi-object spectrographs to efficiently obtain spectra and provide
redshifts for a subset of $\sim 1500$ sources, sufficient to constrain
to $\sim \pm 20$\% the points on AGN luminosity functions (LF) for 0$<$z$<$4.
Currently available instruments ({\it e.g.} the 6.5m
MMT, on which we have 10 nights per year) limit to V\lax22 
but we plan to expand
our sample beyond this once the new generation of
large telescopes come on-line.

\begin{figure}[h]
\plotfiddle{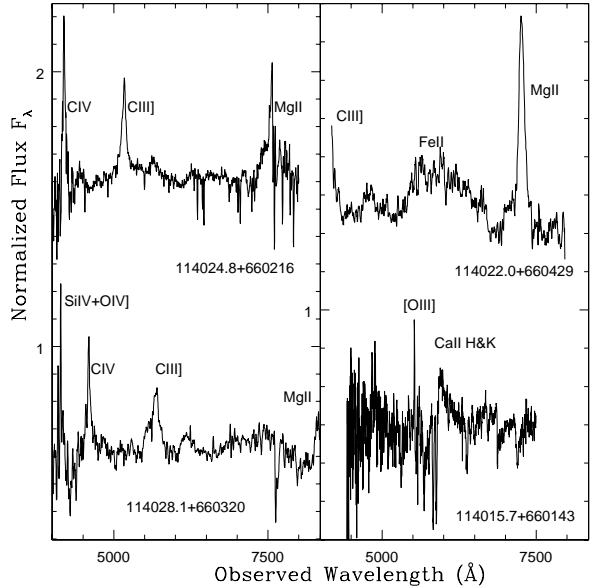}{2.5in}{0}{42}{42}{-200}{-100}
\vskip -2.0in
\hskip 3.0in
\begin{minipage}{2.0in}
\caption{\it  Spectra of 3 ChaMP QSOs and 1 galaxy from Figure~2
taken April 29, 2000 on the
Keck with LRIS, courtesy of the Keck Director.
}
\end{minipage}
\vskip 1.0in
\label{spectra}
\end{figure}

Initial spectra were obtained the Low-Resolution Imaging Spectrograph (LRIS)
on the Keck telescope in director's discretionary time, 29 April 2000.
Slitlets were placed at the positions
of the closest optical couterparts to 12 X-ray soruces in a region of
the MS1137.7$+$6625 field several of which had a low probability due to
the large separation. The resulting 3600sec spectra
revealed 3 stars, 5 QSOs, 2 early-type galaxies,
a narrow emission line galaxy (NELG), and one mystery object
with very smooth continuum (star/BL Lac?).
Four of the spectra: 3 QSOs and 1 galaxy are presented in Figure~4.

{\it This work was supported by NASA contract: NAS8-39073 (CXC)}

 \end{document}